\documentclass[letterpaper, 10 pt, conference]{ieeeconf}
\IEEEoverridecommandlockouts
\usepackage{graphics} 
\usepackage{epsfig} 
\usepackage{mathptmx} 
\usepackage{times} 
\usepackage{amsmath} 
\usepackage{amssymb}  
\usepackage{mathtools}  
\usepackage{cite}
\usepackage{tikz}
\usepackage{subcaption}
\usepackage{booktabs}                     
\usepackage{microtype}                  
\makeatletter
\let\NAT@parse\undefined
\makeatother
\usepackage[breaklinks, colorlinks=true, allcolors=gray, urlcolor=black, linktocpage=true]{hyperref} 

\newif\ifjournal
\journaltrue
\newif\ifpreprint
\preprinttrue \journalfalse

\newcommand{\numberset}{\mathbb}	
\newcommand{\N}{\numberset{N}}
\DeclarePairedDelimiter{\abs}{\lvert}{\rvert} 
\DeclarePairedDelimiter{\norm}{\lVert}{\rVert} 
\DeclarePairedDelimiter\ceil{\lceil}{\rceil}
\renewcommand{\epsilon}{\varepsilon}	

\title{\LARGE \bf
Feedback control of plant-soil autotoxicity via pulse-width modulation
}
\author{Tancredi Rino$^{1}$, Francesco Giannino$^{2}$, 
Davide Fiore$^{3}$
\thanks{$^{1}$Tancredi Rino is with SSM - School for Advanced Studies, Naples, Italy.
        {\tt\small tancredirino@hotmail.it}}%
\thanks{$^{2}$Francesco Giannino is with the Department of Agricultural Sciences, University of Naples Federico II, 80055 Portici, Italy. 
        {\tt\small francesco.giannino@unina.it}}%
\thanks{$^{3}$Davide Fiore is with the Department of Mathematics and Applications ``R. Caccioppoli", University of Naples Federico II, Via Cintia, Monte S.Angelo, 80126 Naples, Italy. {\tt\small  davide.fiore@unina.it}}%
}

\begin{document}
\maketitle
\thispagestyle{empty}
\pagestyle{empty}

\begin{abstract}
Plant–soil negative feedback (PSNF) is the rise in soil of negative conditions for plant performance induced by the plants themselves, limiting the full potential yield and thus representing a loss for the agricultural industry. It has been recently shown that detrimental effects the PSNF has on the growth of plant’s biomass can be mitigated by periodically intervening on the plant/soil system, for example by washing the soil. The periodic control inputs were computed by using an average model of the system and then applied in open-loop. In this paper we present two feedback control strategies, namely a PI and a MPC-based controllers, that, by adapting online the duty-cycle of the periodic control input, guarantee precise regulation of the biomass yield and at the same time robustness to unavoidable modeling errors and perturbations acting on the system. The performance of the proposed control strategies is then validated by means of extensive numerical simulations.
\end{abstract}

\section{Introduction}
\begin{figure}[t]
\centering
\includegraphics[width=.9\columnwidth]{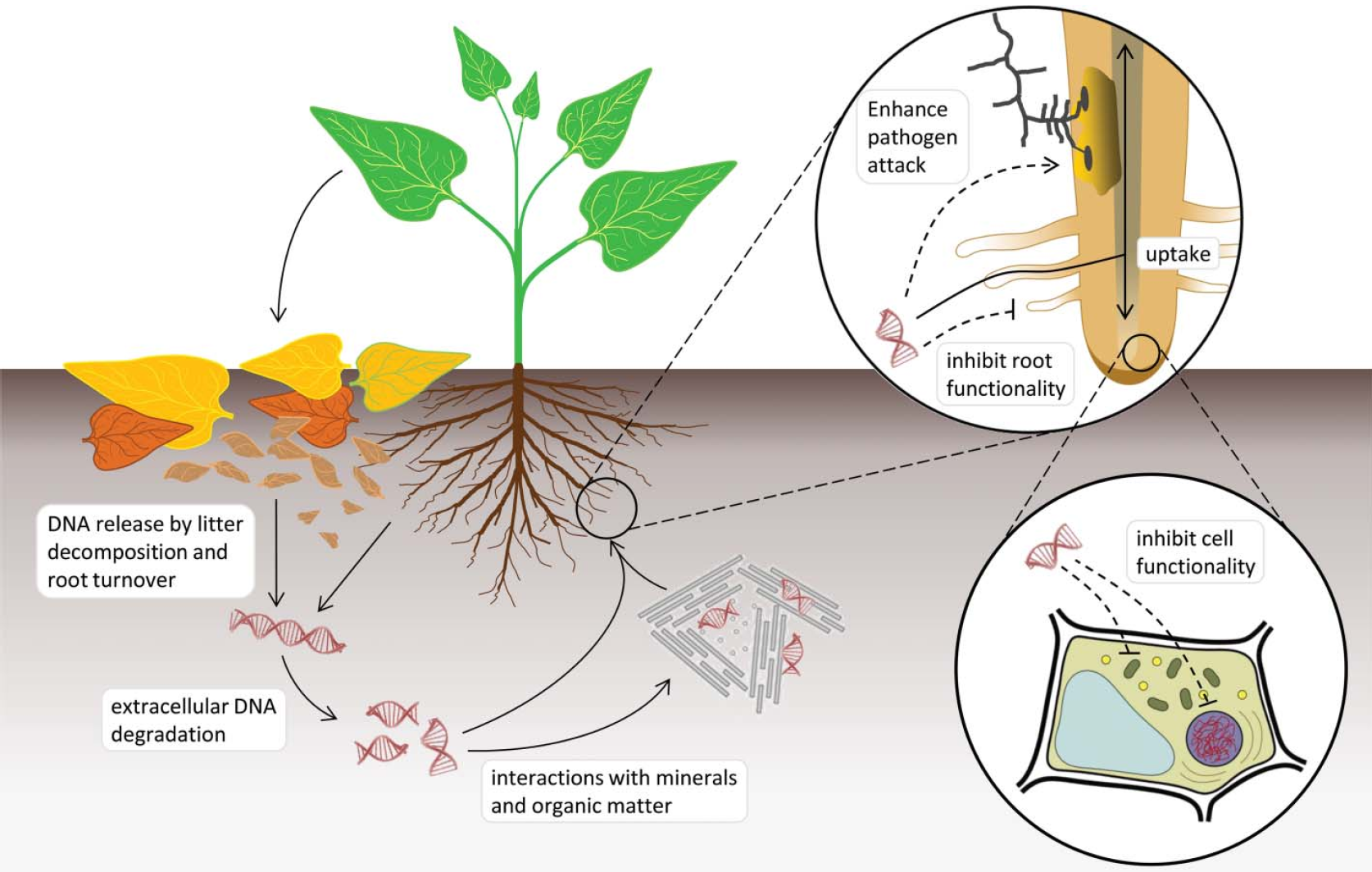}
\caption{Schematic representation of self-DNA soil dynamics
and interactions with plant functionality (reproduced from~\cite{carteni2012negative}).}
\label{subfig:PSNF_DNA}
\end{figure}

The relationship between plants and the soil in which they grow is complex and crucial to agricultural productivity. Plant-soil negative feedback (PSNF) occurs when plants create unfavorable conditions in the soil, limiting their own growth and reducing yields. 
The recognized mechanisms that could produce PSNF are soil nutrient depletion~\cite{ehrenfeld2005feedback}, the accumulation of soil-borne pathogen populations~\cite{packer2000soil}, and the changing composition of soil microbial communities~\cite{klironomos2002feedback}. Moreover, the key mechanism proposed in literature is the development of extracellular self-DNA in the soil caused by plants themselves, as shown in Fig.~\ref{subfig:PSNF_DNA}~\cite{mazzoleni2015inhibitory}.
Plants that accumulate a large amount of leaves, rhizomes, and root litter in the clones' dieback zone, such as a herbaceous plant, negatively affect conspecific regeneration. Indeed, the establishment of clonal plants starts with clusters of densely packed ramets, but, as new ramets grow up in a centrifugal manner, their concentration within the patch gradually decreases, associated by the aging and withering of older shoots, eventually resulting in the formation of a ring-shaped zone~\cite{carteni2012negative}.
Moreover, it has been shown that PSNF also has great effects on agricultural systems~\cite{MARIOTTE2018129}.

To maximize crop production in a robust and reliable manner it is necessary to develop some external control strategy, similarly as what has been done in system biology \cite{milias2016automated,kumar2022platforms,perrino2021automatic}. 
In the context of plant physiology, a first open-loop strategy was recently proposed to mitigate the impact of PSNF on plant growth~\cite{karagiannis2023periodic}, by adding a human periodic intervention on the system, for example by washing the soil or reduce litter from time to time. Indeed, given the periodic nature of the solution, it was possible to study the average model of the system to set the parameters of the control input so to achieve the desired output. Nevertheless, the inherent characteristics of the controller make it susceptible to disturbances coming from various sources, such as noise in the measured data or uncertainty in the model parameters.

In this paper, we introduce two closed-loop control strategies: one based on a Proportional-Integral (PI) algorithm and the other based on Model Predictive Control (MPC).
Inspired by the work presented in~\cite{guarino2020balancing,guarino2019silico}, by adapting in real-time the duty-cycle of the pulsatile inputs, these strategies ensure both performance and robustness of the system.
We validate the performance and the robustness of our proposed strategies by conducting extensive \textit{in silico} experiments that simulate parametric variation on the model parameters, asserting their feasibility and effectiveness in real-world agricultural scenarios.
Finally, we conducted a comparative analysis between the control strategies, discussing their trade-offs. Each strategy offers distinct advantages, and by analyzing their strengths and weaknesses, we provide valuable insights for decision-making and implementation within the agricultural industry.

\section{Mathematical Model}
The biomass/toxin system can be modeled using the following ODEs~\cite{carteni2012negative}:
\begin{equation}
\label{sys:PSNF}
\begin{cases}
    \dfrac{\text{d}B}{\text{d}t}=gB\Bigl(1-\dfrac{B}{B_{\textup{max}}}\Bigr)-dB-sBT \\[2ex]
    \dfrac{\text{d}T}{\text{d}t}=c(dB+sBT)-kT
\end{cases},
\end{equation}
where the state variables $B$ and $T$ denote the superficial density (in $\mathrm{kg \cdot cm^{-2}}$) of the plant biomass and the toxic compounds, respectively. All the parameters in~\eqref{sys:PSNF} are positive real and constant. Specifically, $g$ is the growth rate of the plant in absence of toxins, $B_{\textup{max}}$ is the carrying capacity of the field, $d$ is the natural death rate of the plant, $s$ is the plant sensitivity to the presence of the toxin, $c$ is the toxin production rate, and $k$ is the toxin decay rate. Their units of measure are reported in Table~\ref{tab:parameter}. In this model the plant biomass $B$ is assumed to follow a logistic growth dynamics. However, the carrying capacity $B_{\textup{max}}$ cannot be reached due to the mortality of the plant, and to the detrimental effect of the toxicity induced by vegetative propagation, which is described by the further term $- s B T$. Instead, the rate of production of the toxin compounds $T$ is proportional to the total decay rate of the biomass, and it degrades with a first-order dynamics at rate $k$.

The system~\eqref{sys:PSNF} has two physically admissible equilibria~\cite{carteni2012negative}. With the numerical values of the parameters reported in Tab.~\ref{tab:parameter}, these equilibria are the unstable origin $(0, 0)$ and the stable point
\begin{equation}\label{eq:equilibrium_Bstar-Tstar}
    (B^*, T^*) = \biggl( \dfrac{1 + \beta - \Gamma}{2\beta}, \dfrac{\beta - 1 - 2\alpha\beta + \Gamma}{2\alpha\beta^2} \biggr) ,
\end{equation}
where $\alpha = d / g$, $\beta = c s B_{\textup{max}} / k$, and $\Gamma = \sqrt{(\beta - 1)^2 + 4\alpha\beta}$.

\medskip
The production of the plant biomass can be increased by decreasing the level of toxic compounds in the soil. 
This is possible by, for example, either washing or burning the soil.
This human intervention can be modeled as an external control action $u$, that is
\begin{equation}\label{sys:PSNF_with_u}
\begin{cases}
    \dfrac{\text{d}B}{\text{d}t}=gB\Bigl(1-\dfrac{B}{B_{\textup{max}}}\Bigr)-dB-sBT \\[2ex]
    \dfrac{\text{d}T}{\text{d}t}=c(dB+sBT)-kT - u(t, T)
\end{cases},
\end{equation}
where $u(t, T) \ge 0$, the removal rate of the toxic compound, is proportional to its current density in the soil $T$.
In order to provide to human operators (e.g., gardeners) with the simplest yet effective control ‘recipe’ to follow, the control input $u(t, T)$ was assumed to be periodic and discontinuous \cite{karagiannis2023periodic}, namely a pulse wave of period $P$, duty-cycle $D\in[0, 1]$, and constant amplitude $\gamma > 0$, that is,
\begin{equation}\label{eq:u_openloop}
    u(t, T) = \gamma \, \textup{s}_{\textup{q}}(t/P) \, T,
\end{equation}
where $s_q(t)$ is the pulse wave taking values 0 and 1, with period 1, duty-cycle $D$, and the amplitude $\gamma$ depends on the specific choice of physical control action (e.g., washing or burning) and other physical properties of the system.

\begin{table}[tp]
    \centering
    \caption{Model parameters and their assigned values~\cite{carteni2012negative}.}
    \begin{tabular}{ccc}
    \toprule
    Parameter & Unit & Assigned value \\
    \midrule
    $g$ & $\mathrm{month^{-1}}$ & $0.5$ \\
    $B_{\textup{max}}$ & $\mathrm{kg \, cm^{-2}}$ & $1$ \\
    $d$ & $\mathrm{month^{-1}}$ & $0.015$ \\
    $s$ & $\mathrm{cm^2\,kg^{-1}\,month^{-1}}$ & $0.15$ \\
    $c$ & - & $0.5$ \\
    $k$ & $\mathrm{month^{-1}}$ & $0.05$ \\
    \bottomrule
    \end{tabular}
    \label{tab:parameter}
\end{table}

The model~\eqref{sys:PSNF_with_u} is a nonlinear system subject to periodic forcing $u(t, T)$. 
Hence, by means of the averaging method \cite{khalil2002nonlinear,fiore2018analysis}, it is possible to obtain an average model that describes the evolution of the mean value of $B$ and $T$ over a period $P$.
In what follows we denote these quantities with $B_{\textup{av}}$ and $T_{\textup{av}}$ (see Appendix~\ref{sec:appendix_ave_model} for further details).

By exploiting the average model, it was shown in \cite{karagiannis2023periodic} that it can be derived a relationship between the parameters of the pulse-width modulated (PWM) input signal $u(t, T)$ in \eqref{eq:u_openloop} and $B_\mathrm{av}$, so that the latter is asymptotically bounded. 
More precisely, by defining a bound $\delta > 0$, the \emph{control goal} was to design $u(t, T)$ so that
\begin{equation}
    \label{eq:controlGoal_general}
    \limsup_{t \to \infty}\,\norm*{B_\mathrm{av}(t) - \hat{B}} < \delta,
\end{equation}
where $\hat{B}=\frac{B_{\textup{max}}}{g} (g - d)$ is the value of biomass if it were possible to completely remove the toxin from the soil (i.e. $T^*=0$). However, this value is not physically achievable~\cite{karagiannis2023periodic}.
From the above mentioned relationship, $B_\mathrm{av}(\gamma\, D)$, and the control goal \eqref{eq:controlGoal_general} one gets the following condition on the parameters in \eqref{eq:u_openloop}
\begin{equation}
    \label{eq:controlGoal_PWM}
    \gamma D > - \delta c s B_{\textup{max}} - k + \dfrac{c s B_{\textup{max}}^2 (g - d)}{g} .
\end{equation}
However, due to the presence of unavoidable disturbances and uncertainties in the model, the open-loop design in \eqref{eq:controlGoal_PWM} cannot guarantee robustness and reliable performance in realistic scenarios.

\section{Closed-loop Control}
\begin{figure}[tp]
\centering
\includegraphics[width=\columnwidth]{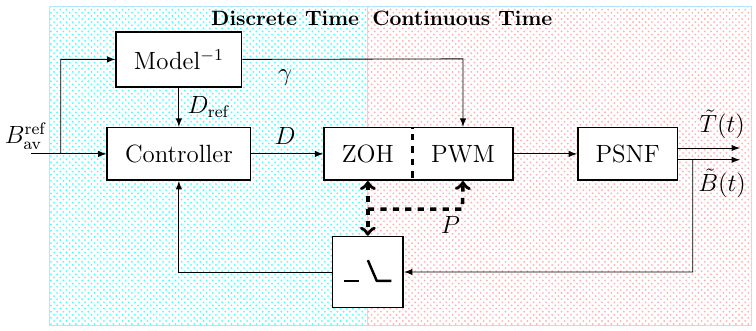}
\caption{Block diagram of the proposed closed-loop control strategies.
The feedforward model-based inversion block evaluates the reference duty-cycle $D_{\textup{ref}}$ given the desired average value of the biomass  $B^{\textup{ref}}_{\textup{av}}$.
At each sampling period, starting from the previously computed $D_\mathrm{ref}$, the feedback control evaluates the duty-cycle $D$ that minimize the error between the desired reference value $B_\mathrm{av}^\mathrm{ref}$ and the measured mean value of $B$ over the period $P$. 
Note that the control system evolves in discrete time (with sampling time $P$) while the controlled system in continuous time.}
\label{fig:feedback_PSNF_scheme}
\end{figure}

\begin{figure*}[t]
\centering
\subfloat[][\label{fig:feedback_PSNF_scheme_PI}]
{\includegraphics[width=.95\columnwidth]{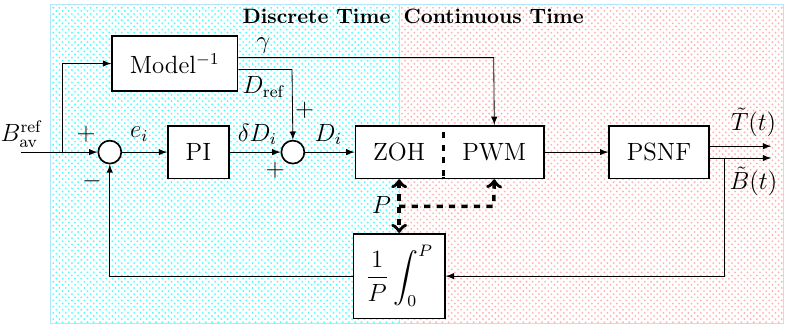}} \quad
\subfloat[][\label{fig:feedback_PSNF_scheme_MPC}]
{\includegraphics[width=.95\columnwidth]{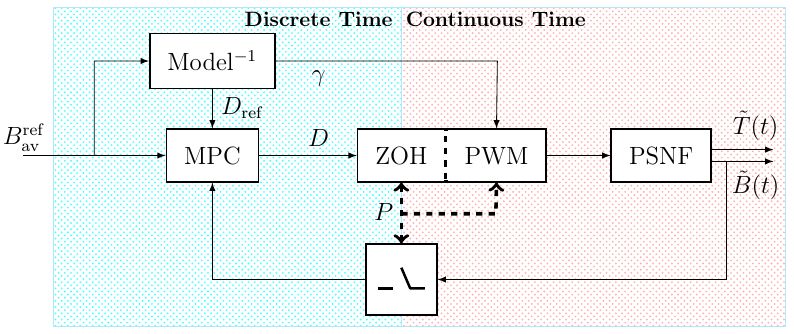}}
\caption{Block diagrams of proposed feedback control strategies. (a) PI/PWM control strategy. Given the set-point for the average biomass $B_{\textup{ref}}$, two actions regulate the parameters of the PWM inputs that feed the system. 
The feedforward action consists in a model-based inversion that evaluates the duty-cycle $D_{\textup{ref}}$. A proportional-integral controller composes the feedback loop. At each time period $t_i = iP$, the error $e_i$ is corrected by a PI controller that evaluates the correction $\delta D_i$ to $D_{\textup{ref}}$. 
(b) MPC control strategy. At each discrete time step $i$, the MPC finds the sequence of duty-cycles $\{D^i_0, D^i_1, \dots, D^i_{N_H - 1}\}$ that minimizes the cost function $J^i$ over the prediction horizon $t_H = N_H P$. Then, only the first element of the sequence is selected ($D^i = D_0$), and the corresponding pulsatile control signal is applied to the system in the time interval $[t_i, t_i + P]$.}
\end{figure*}

To solve the problem described above, we propose a feedback control approach that adapts and changes online the duty-cycle $D$. The strategy is based on two control actions, as shown in Fig.~\ref{fig:feedback_PSNF_scheme}:
\begin{itemize}
    \item a feedforward action, taken from~\cite{karagiannis2023periodic}, that precomputes the value of the duty-cycle $D_{\textup{ref}}$ required to achieve the control goal in nominal conditions, i.e. in the absence of any perturbation (this is done by inverting the average model, reported in Appendix \ref{sec:appendix_ave_model});
    \item a feedback action that corrects the duty-cycle online.
    Specifically, we implemented two feedback control algorithms:
    \begin{itemize}
        \item a Proportional-Integral (PI/PWM) controller that updates the duty-cycle based on present and past control errors 
        (see Fig.~\ref{fig:feedback_PSNF_scheme_PI});
        \item a Model Predictive Controller (MPC) that evaluates the duty-cycle by optimizing a desired cost function at each sampling period (see Fig.~\ref{fig:feedback_PSNF_scheme_MPC}).
    \end{itemize}
\end{itemize}
In the following we consider a reference value of the average biomass $B_{\mathrm{av}}^{\mathrm{ref}}$ equal to $0.9$. 
This value was arbitrarily chosen to be greater than the value achievable without any control action, $B^*$ in~\eqref{eq:equilibrium_Bstar-Tstar}, and less than the ideal, unachievable, value $\hat{B}$ in~\eqref{eq:controlGoal_general}.
Moreover, in scenarios in which in the same environment, more than one biomass is present, keeping their density lower than their carrying capacity $B_{\textup{max}}$ can be help co-existence and better use of resources. 

\subsection{Feedforward action.}
\begin{figure}[tp]
\centering
\includegraphics[width=.8\columnwidth]{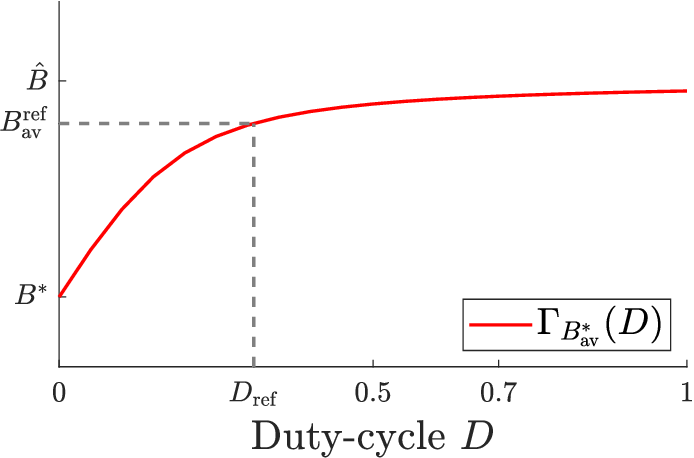}
\caption[Feedforward action]{
Functional relationship between the average value of the biomass at steady state, $B_\mathrm{av}^*$ and the duty-cycle $D$ of the periodic control input $u(t, T)$.
Given an initial reference value of biomass $B^{\textup{ref}}_{\textup{av}}$ on the $y$-axis, the $D_{\textup{ref}}$ is found as the corresponding $x$-axis of $\Gamma_{B^*_{\textup{av}}}(D)$. Here $B^{\textup{ref}}_{\textup{av}} = 0.9$ from which we get $D_{\textup{ref}} \approx 0.31$.}
\label{fig:feedforward_Dref}
\end{figure}

The feedforward block exploits the relationship between the features of the periodic input $u(t, T)$ in~\eqref{eq:u_openloop} and the response of system~\eqref{sys:PSNF_with_u} when subjected to it.
Specifically, by means of the averaging theory, the value of the equilibrium point of the average model $B_\mathrm{av}^*$ (see equation~\eqref{sys:PSNF_averaged} in Appendix) can be derived as a function of the duty-cycle $D$. 
Hence, by inverting $B_\mathrm{av}^*(D)$ it is possible to compute the value of the duty-cycle $D_\mathrm{ref}$ corresponding to %some desired
$B_\mathrm{av}^\mathrm{ref}$, that is, $D_\mathrm{ref}=B_\mathrm{av}^{*^{-1}}(B_\mathrm{av}^\mathrm{ref})$.
This value can be computed either by inverting each time the above function or by querying a pre-computed curve $\Gamma_{B^*_{\textup{av}}}(D)$ obtained by interpolation (see Fig.~\ref{fig:feedforward_Dref}).

\subsection{PI feedback control strategy.}
The PI/PWM controller (Fig.~\ref{fig:feedback_PSNF_scheme_PI}) evaluates at every sampling time instant $t_i = i P$ for $i \in \N$, the duty-cycle $D_i$, by adding a correction $\delta D_i$ to the reference value $D_{\textup{ref}}$ returned by the feedforward block. 
That is, the duty-cycle is updated online as 
\[
D_i = D_{\textup{ref}} + \delta D_i \text{,}\quad i\in \N ,
\]
where $\delta D_i$ is computed as
\[
\delta D_i = K_P\,e_i + K_I \sum_{j = 0}^i e_j ,
\]
with initial condition $D_0 = D_{\textup{ref}}$. 
$K_P$ and $K_I$ are the gains of the PI controller and $e_i$ is the error between the reference value~$B^{\textup{ref}}_{\textup{av}}$ and the current measured mean value of the biomass~$B$ over a time period~$P$, that is, 
\[
e_i = B^{\textup{ref}}_{\textup{av}} - \frac{1}{P} \int_{(i - 1)P}^{i P} B(\tau)\,d\tau .
\]

The tuning of the PI gains was carried out by performing numerical simulations in \texttt{MATLAB}, evaluating several metrics.
Specifically, we selected 1500 pairs of gain values $K_P$ and $K_I$ sampled uniformly form the intervals $K_P \in [0, 2]$ and $K_I \in [1, 30]$, with steps 0.1 and 1, respectively.
For each pair $(K_P,K_I)$ the closed-loop system was simulated for $N=20$ time periods.
The pairs $(K_P, K_I)$ corresponding to saturated values of the duty-cycle ($D\approx 1$ or $0$) were removed from the analysis. 
Likewise, pairs for which the duty-cycle did not reach the steady state within $7$ periods. 
The above mentioned metrics are:
\begin{itemize}
    \item The \textit{average relative percentage steady-state error}, defined as
    \begin{equation}\label{eq:e_rperc}
        e_{r\%} = \abs*{\frac{B^{(5)}_{\textup{mean}} - B^{\textup{ref}}_{\textup{av}}}{B^{\textup{ref}}_{\textup{av}}}}\times100\% ,
    \end{equation}
    where $B^{(5)}_{\textup{mean}}$ is the value of $B(t)$, averaged over the last 5 periods.
    \item The settling time $T_{s_{10\%}}$ of $B(t)$, computed as the number of periods $P_s$ needed to reach the steady state, that is, 
    \begin{equation}\label{eq:metrics_set-time}
        P_s = \ceil*{\frac{T_{s10\%}}{P}} .
    \end{equation}
    \item The maximum value of $D$ reached during the simulation, that is,
    \begin{equation}
        D_{\textup{max}} = \max_{i \in [1, N]} D_i ;
    \end{equation}
    \item \textit{Integral Square Error} (ISE), defined as
    \begin{equation}\label{eq:ISE}
        \text{ISE} = \int_{t_0}^{t_f} e_{\textup{MA}}(\tau)^2\,d\tau ,
    \end{equation}
    where $t_0 = P$ and $t_f = N P$ is the time instant at the end of the experiments, and $e_{\textup{MA}}(t)$ is the relative error as
    \[
        e_{\textup{MA}}(t) = \dfrac{B_{\textup{MA}}(t) - B^{\textup{ref}}_{\textup{av}}}{B^{\textup{ref}}_{\textup{av}}},
    \]
    in which $B_{\textup{MA}}(t)$ is  the moving average of $B(t)$ with width equal to $P$, that is
    \begin{equation}\label{eq:B_MA}
        B_{\textup{MA}}(t) = \frac{1}{P} \int_{t - P}^t B(\tau)\,d\tau\text{, \quad for $t \ge P$.}
    \end{equation}
    By cumulatively calculating the squared errors over a period, the ISE penalizes large errors.
    \item \textit{Integral Time-weighted Absolute Error} (ITAE), defined as
    \begin{equation}\label{eq:ITAE}
        \text{ITAE} = \int_{t_0}^{t_f} \tau\abs{e_{\textup{MA}}(\tau)}\,d\tau .
    \end{equation}
    Notice that the ITAE penalizes more small persistent errors that occur at steady state than large errors at the beginning of the experiments. 
\end{itemize}

\begin{figure}[tp]
\centering
\includegraphics[width=\columnwidth]{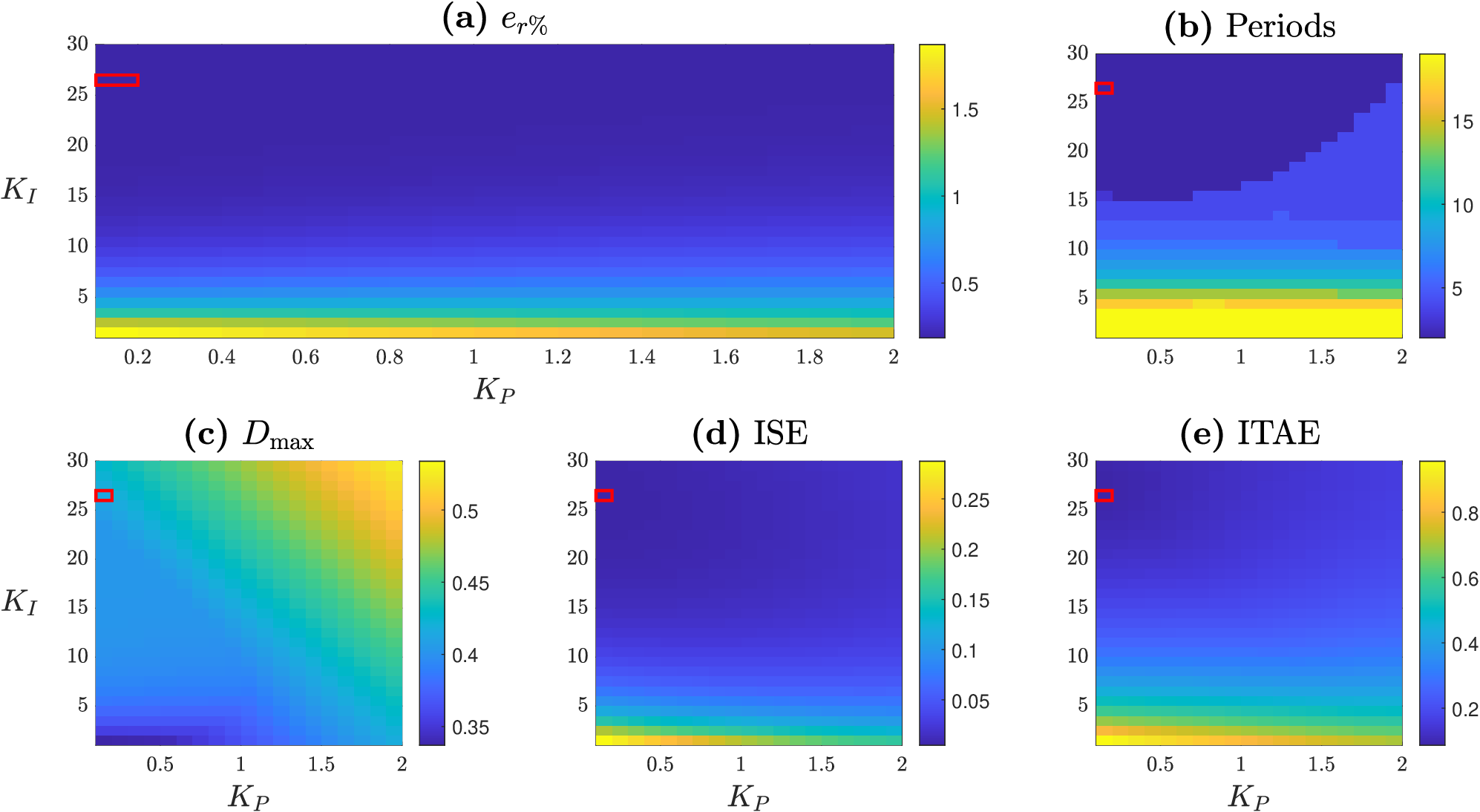}
\caption[Tuning of the PI controller]{Tuning of the PI controller. The red box indicates the values of PI gains $K_P = 0.1$, $K_I = 26$ that were selected as those giving the best compromise between all the metrics we considered, and used for control experiments.}
\label{fig:tunin_PI}
\end{figure}

\begin{figure*}[tp]
\centering
\subfloat[][\label{subfig:PI_0_9}]{\includegraphics[width=.88\columnwidth]{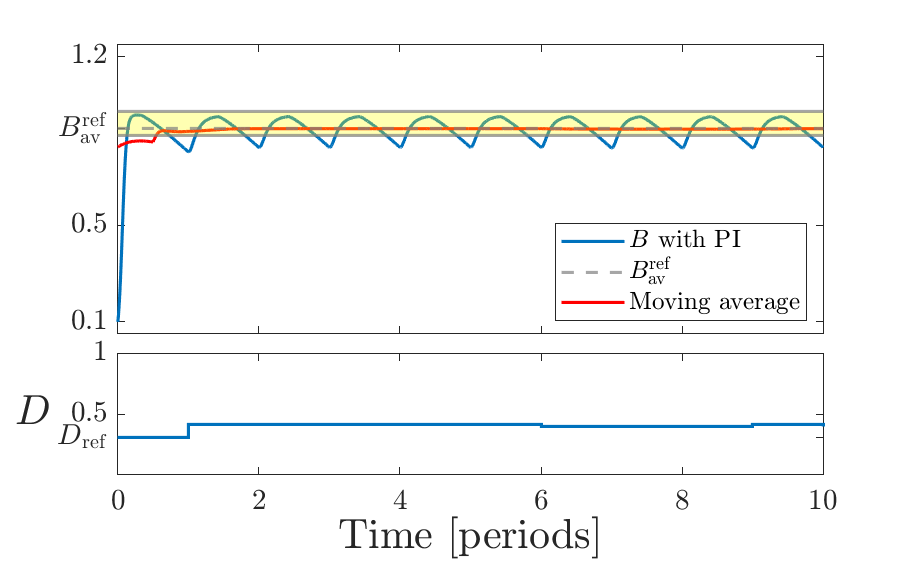}} \quad
\subfloat[][\label{subfig:GA_0_9}]{\includegraphics[width=.88\columnwidth]{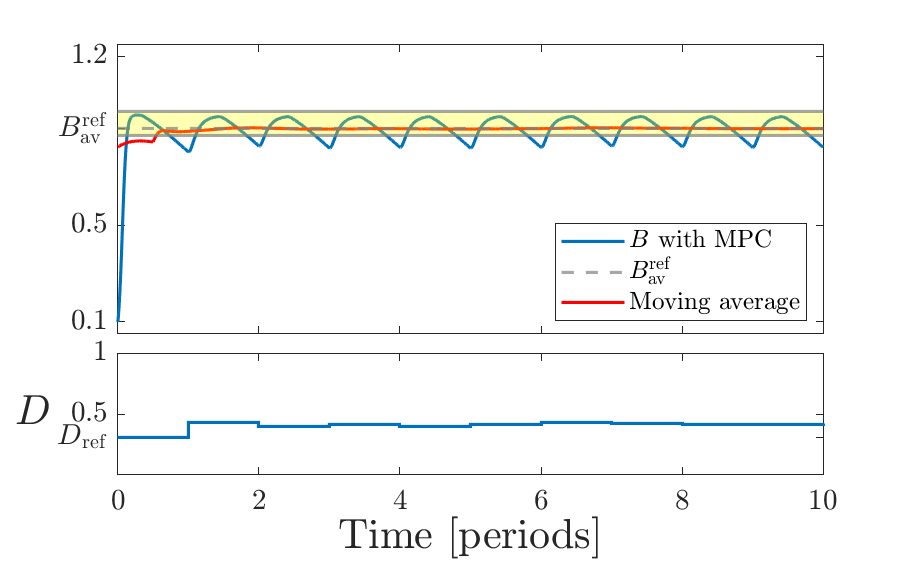}}
\caption[System time evolution]{Time evolution of system in closed-loop, with $B^{\textup{ref}}_{\textup{av}} = 0.9$ (dashed line), and bound $\delta = 0.1$ (in yellow).
Top panels show the time evolution of biomass $B(t)$ compared with their moving averages $B_\mathrm{MA}(t)$ defined in~\eqref{eq:B_MA}, while the bottom panels show the time evolution of duty-cycle. (a) PI controller strategy with gains $K_P = 0.1$ and $K_I = 26$. (b) MPC with a genetic algorithm to generate the sequences of duty-cycles $\{D^i_0, D^i_1, \dots, D^i_{N_H - 1}\}$.}
\label{fig:B(t)}
\end{figure*}

The result of the tuning of the control gains is reported in Figure~\ref{fig:tunin_PI}. 
The values of $K_P = 0.1$ and $K_I = 26$ were selected  
as the pair corresponding to the minimum value of percentage steady-state error, ISE and settling time metrics. The other metrics have been used only for validation of the performance.
In Fig.~\ref{subfig:PI_0_9} it is reported the time evolution of biomass $B$ and duty-cycle $D$ when the system is controlled by the PI/PWM previously tuned.
After two periods, the mean value of the biomass (represented by its moving average as a red line) is asymptotically regulated to the reference value, with a small offset due to numerical quantization of the control parameter~$D$.

\subsection{MPC feedback control strategy.}
The MPC (Fig.~\ref{fig:feedback_PSNF_scheme_MPC}) chooses, at each sampling time $t_i = iP$, the duty-cycle $D^i$ to be applied in the next control cycle (of duration $P$), by solving an online optimization problem on a finite prediction horizon interval of length $t_H = N_H P$, with $N_H = 5$. 
The value of $N_H$ was selected heuristically such that computation time is not a problem and the system has reached steady-state within $N_H$ periods. 
Specifically, at each step $i$, the algorithm finds the sequence of duty-cycles $\{D^i_0, D^i_1, \dots, D^i_{N_H - 1}\}$ of length $N_H$ that minimizes the cost function $J^i$, defined as
\begin{equation}
    \label{eq:costFunction}
    J^i(\{D^i_0, \dots, D^i_{N_H -1}\}) = \sum_{j = 0}^{N_H - 1}
    e_r(t, D^i_j) 
    ,
\end{equation}
where $e_r(t, D^i_j)$ is the relative error defined as 
\[
e_r(t, D^i_j) = \abs*{\frac{B_{\mathrm{mean}}(t, D^i_j) - B^{\textup{ref}}_{\textup{av}}}{B^{\textup{ref}}_{\textup{av}}}} ,
\]
and $B_{\mathrm{mean}}(t, D^i_j)$ is the mean value of the biomass evaluated over a period~$P$, that is
\[
B_{\mathrm{mean}}(t, D^i_j) = \frac{1}{P} \int_{(i + j)P}^{(i + j + 1)P} B(\tau, D^i_j)\,d\tau .
\]

We numerically solve the optimization problem to find the sequences of duty-cycles $\{D^i_0, \dots, D^i_{N_H - 1}\}$ using a genetic algorithm~\cite{banzhaf1998genetic}, that relies on the concepts of genetics and natural selection for conducting search-based operations. 
Specifically, at each time step $i$, we initialize the population to 20 randomly chosen individuals. The values $D^i_1$ are drawn with uniform probability from the interval $\left[- \frac{D^{i - 1}}{2},\, \frac{D^{i - 1}}{2}\right]$, with $D^0 = D_{\textup{ref}}$. 
The other values of the population $D^i_{j + 1}$, for $j \in [0, N_H - 2]$, are chosen as small perturbation of~$D^i_0$, specifically, they are drawn from a normal distribution centered at $D^i_0$ with standard deviation $0.05$ (trimming values outside the definition interval $[0,1]$). 
Then, the parents for crossover are chosen with tournament selection, with a tournament size $K = 4$. With probability $p_c = 0.5$ we perform the crossover, by taking the average of the parents' values. Subsequently, the offspring are mutated with probability $p_m = 0.1$ using a Gaussian mutation. We repeated these steps 50 times, unless an additional stopping criteria based on a maximum number of generations of stalls equal to 10.
 
The time evolution of biomass $B$ and of the duty-cycle $D$ when the system is controlled with the MPC is reported in Fig.~\ref{subfig:GA_0_9}.
Similarly as with the PI/PWM, after two periods, the mean value of the biomass (represented by its moving average as a red line) settles to a steady-state value close (around $0.02\%$) to the desired reference value $B^{\textup{ref}}_{\textup{av}} = 0.9$.

\subsection{Robustness of the controllers.}
\begin{figure*}[tp]
\centering
\includegraphics[width=.84\textwidth]{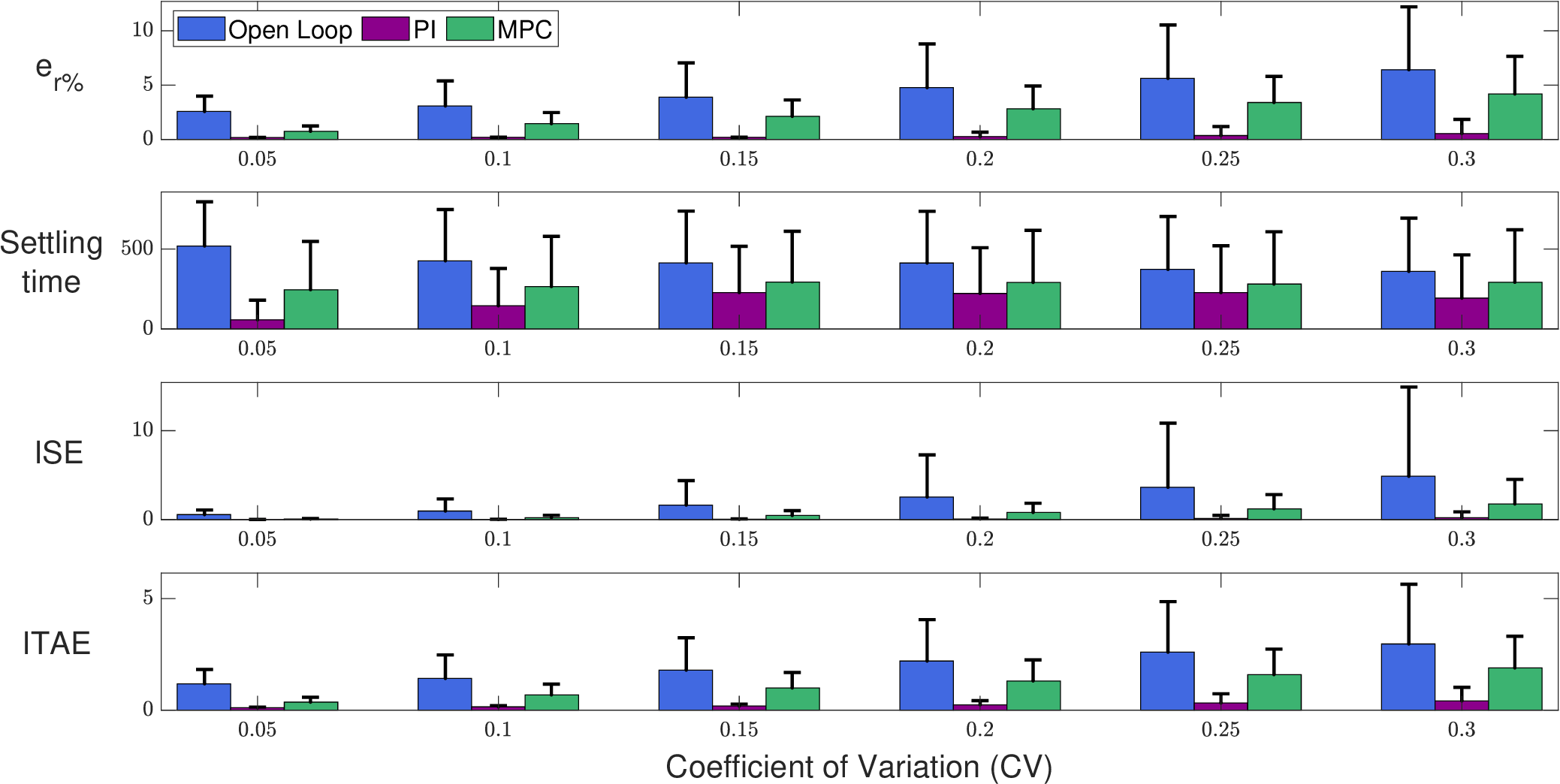}
\caption[Sensitivity to parameter variations]{Sensitivity to parameter variations: mean and standard deviation of the percentage of error at steady state, settling time, ISE and ITAE as the parameters' variability increases. Blue bars are obtained with the open-loop controller from~\cite{karagiannis2023periodic}, purple bars are obtained using the PI/PWM controller, and green bars are obtained using the MPC. 
For each value of $\textup{CV} \in\{0.05, 0.1, 0.15, 0.2, 0.25, 0.3\}$ we performed $n = 50$ simulations drawing independently the parameters from a normal distribution centered at their nominal value $\hat{\rho}$ with standard deviation $\sigma = \textup{CV} \cdot \hat{\rho}$. The reference signal was set to $B^{\textup{ref}}_{\textup{av}} = 0.9$, and the PI gains were chosen as $K_P = 0.1$ and $K_I = 26$.}
\label{fig:CV}
\end{figure*}

We tested the robustness of both closed-loop strategies by introducing parametric variations in the parameters of Table~\ref{tab:parameter}, similarly as done in~\cite{martinelli2023multicellular}.
We modeled this effect by drawing each parameter, say $\rho$, from a normal distribution centered on its nominal value $\hat{\rho}$ with standard deviation $\sigma = \textup{CV} \cdot \hat{\rho}$, where $\textup{CV}$ is the coefficient of variation. 
We compared robustness, as $\textup{CV}$ increases in the set $\{ 0.05,\, 0.1,\, 0.15,\, 0.2,\, 0.25,\, 0.3 \}$, by evaluating the average relative percentage steady-state error, ISE, ITAE, and the settling time, averaged over $n = 50$ experiments.
The results of the above comparison are reported in Fig.~\ref{fig:CV}. 
The transient performance, represented by the settling time in the figure, deteriorates in both cases.
However, the PI controller showed better robustness then the MPC at steady state due to its embedded integral action.

\section{Discussion and Future Works}
To compensate for inevitable disturbances or parametric discrepancies in real-world scenarios, we implemented two control algorithms, namely a PI/PWM and an MPC. These algorithms were introduced to correct the values predicted by the model inversion block and ensure accurate regulation of the output. 
The controllers showed to be robust in simulations, maintaining high performance despite uncertainties in parameters and demonstrating minimal sensitivity to the precise choices of controller gains.
The choice between the two strategies depends on specific industrial requirements. 
If accurate regulation of the biomass is more important, then employing a PI controller would be more suitable.
On the other hand, the MPC could easily take into account additional control requirements, not considered at this stage (further discussed later).
These controllers, designed for the enhancement of biomass yield, have the potential for immediate application in agricultural systems. In fact, they can be used effectively to optimize treatments in situations where the impact of PSNF is not negligible. 

\begin{figure}[tp]
\centering
\includegraphics[width=.95\columnwidth]{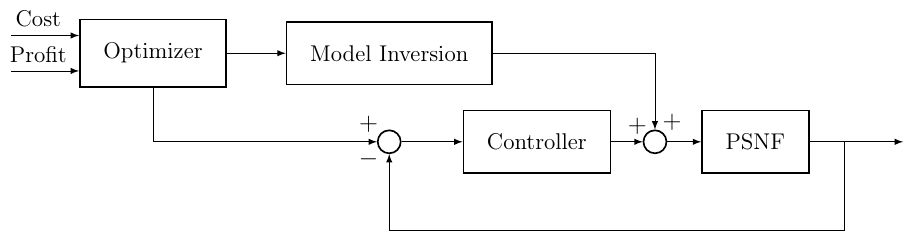}
\caption[Possible future work]{A possible implementation of a feedback control strategy to robustly regulate the biomass produced, given cost and
profit constraints.}
\label{fig:futureWork}
\end{figure}

A further improvement of the control strategies we present here can consider additional constraints in the evaluation of the control action $u(t,T)$, for example constraints on the cost of the control action, use of limited resources (e.g., nutrients, water, light), in order to maximize the profit derived from the biomass production.
The mathematical models developed here can indeed be exploited to build a model inversion block that can be used in combination with the feedback control action (see Fig.~\ref{fig:futureWork}). Specifically, an optimizer computes the reference value for the biomass, given the cost of the control action and the profit deriving from the crop. This value is then fed to a model inversion block that computes the parameters $\gamma$ and $D$ of the pulse wave policy to be applied to the system. Finally, the control algorithm (e.g., the PI, or the MPC or an extremum seeking) can be used to correct the values predicted by the model inversion to compensate for unavoidable disturbances or parametric mismatches in the real world.
From an agronomic perspective, these control algorithms could increase production without introducing additional chemical fertilizers, thus reducing environmental pollution.
Furthermore, the reduction of toxins could have very low costs (both with water and physical removal), entirely compatible with increasing agricultural production.
Notice that any constraint related to the cost of the control action or the profit deriving from the biomass can be easily included in the function cost of the MPC approach.

Future works will be aimed at conducing \textit{in vivo} experiments to validate the proposed control designs, and also to investigate more advanced scenarios in which there are two or more biomasses each producing a toxic compound that can, at the same time, reduce the growth rate of the parent biomass and increase the growth rate of the others.

\appendices
\section{} 
\label{sec:appendix_ave_model}
System~\eqref{sys:PSNF_with_u} can be written in nondimensional form, using the following dimensionless variables~\cite{carteni2012negative}\ifjournal~(see more in~\cite{rino2023feedback})\fi,
\[
\tilde{B}=\frac{B}{B_{\textup{max}}}, \quad \tilde{T}=\frac{k}{cdB_{\textup{max}}}T, \quad \tilde{t}=gt .
\]
\ifpreprint
Therefore, system~\eqref{sys:PSNF_with_u} becomes
\begin{equation}\label{sys:PSNF_dim_with_u}
    \begin{cases}
        \dfrac{\text{d}\tilde{B}}{\text{d}\tilde{t}} = \tilde{B}(1 - \tilde{B}) - \alpha \tilde{B} - \alpha\beta \tilde{B}\tilde{T} \\[2ex]
        \dfrac{\text{d}\tilde{T}}{\text{d}\tilde{t}} = \kappa \tilde{B} + \beta\kappa \tilde{B}\tilde{T} - \kappa \tilde{T} - \gamma\textup{s}_{\textup{q}}(\tilde{t}/P)\tilde{T}
    \end{cases},
\end{equation}
where $\kappa = k/g$. Re-scaling the time as $\tau = \tilde{t}/P$, system~\eqref{sys:PSNF_dim_with_u} can be written as
\begin{equation}\label{sys:PSNF_recast}
    \begin{cases}
        \dfrac{\text{d}\tilde{B}}{\text{d}\tau} = \epsilon\bigl[ \tilde{B}(1 - \tilde{B}) - \alpha \tilde{B} - \alpha\beta \tilde{B}\tilde{T} \bigr] \\[2ex]
        \dfrac{\text{d}\tilde{T}}{\text{d}\tau} = \epsilon\bigl[ \kappa \tilde{B} + \beta\kappa \tilde{B}\tilde{T} - \kappa \tilde{T} - \gamma\textup{s}_{\textup{q}}(\tau)\tilde{T} \bigr]
    \end{cases}
\end{equation}
where $\epsilon = P$. 
\fi
Thus, the dimensionless average model of
\ifpreprint
\eqref{sys:PSNF_recast} is described by
\fi
\ifjournal
\eqref{sys:PSNF_with_u} is described by
\fi
\begin{equation}\label{sys:PSNF_averaged}
    \begin{cases}
        \dfrac{\text{d}\tilde{B}_{\textup{av}}}{\text{d}\tau} = \epsilon\bigl[ \tilde{B}_{\textup{av}}(1 - \tilde{B}_{\textup{av}}) - \alpha \tilde{B}_{\textup{av}} - \alpha\beta \tilde{B}_{\textup{av}}\tilde{T}_{\textup{av}} \bigr] \\[2ex] 
        \dfrac{\text{d}\tilde{T}_{\textup{av}}}{\text{d}\tau} = \epsilon\bigl[ \kappa \tilde{B}_{\textup{av}} + \beta\kappa \tilde{B}_{\textup{av}}\tilde{T}_{\textup{av}} - \kappa \tilde{T}_{\textup{av}} - \gamma D\tilde{T}_{\textup{av}} \bigr]
    \end{cases},
\end{equation}
\ifjournal
where $\epsilon = P$, $\tau = \tilde{t}/P$\ifpreprint~\cite{karagiannis2023periodic}\fi, $\kappa = k/g$, and $\tilde{B}_{\textup{av}}$ and $\tilde{T}_{\textup{av}}$ are, respectively, the average biomass and toxin abundances over a period~$P$.
\fi
\ifpreprint
where $\tilde{B}_{\textup{av}}$ and $\tilde{T}_{\textup{av}}$ are, respectively, the average biomass and toxin abundances over a period~$P$. 
\fi
The first coordinate of the non-trivial positive equilibrium point of the system~\eqref{sys:PSNF_averaged} is
\begin{equation}\label{eq:B*_av}
    \tilde{B}^*_{\textup{av}} = \dfrac{\beta + 1 + \eta - \sqrt{(\beta - 1 - \eta)^2 + 4\alpha\beta(1 + \eta)}}{2\beta},
\end{equation}
where $\eta = \gamma D / \kappa$. For $\alpha \ll 1$, it follows
\begin{equation}\label{eq:approx_B*av(D)}
    \tilde{B}^*_{\textup{av}}(D) \approx \dfrac{1 + \eta}{\beta} = \tilde{B}^* + \frac{\gamma}{\kappa\beta}D.
\end{equation}

\bibliographystyle{IEEEtran}
\bibliography{refs}

\end{document}